\documentclass[5p, preprint]{elsarticle}
\biboptions{sort&compress}
\usepackage{lineno,hyperref,isotope,mathrsfs}
\usepackage{amsmath,bbold,amssymb,epsfig,feynmp,color,ifpdf}
\usepackage{slashed,nicefrac,exscale,multirow,times,txfonts}
\usepackage{epstopdf}

\begin{document}
\begin{frontmatter}

\title{$B(E2)$ Serves as a Robust Signature of $N=32,34$ Shell Evolution}
\author[1,6]{Jia Liu}
\author[4,5]{Yi Fei Niu}
\author[1,6]{Xiao Hua Li\corref{cor1}}
\author[1,6]{Wen Luo}
\author[2,3]{Wen Hui Long\corref{cor1}}
\cortext[cor1]{Corresponding authors.\\[6pt]
Xiao-Hua Li (lixiaohuaphysics@126.com)\\[6pt]
Wen-Hui Long (longwh@lzu.edu.cn)}
\address[1]{School of Nuclear Science and Technology, University of South China, Hengyang 421001, China}
\address[6]{Key Laboratory of Advanced Nuclear Energy Design and Safety, Ministry of Education, Hengyang, 421001, China}
\address[4]{School of Physics and Astronomy, Shanghai Jiao Tong University, Key Laboratory for Particle Astrophysics and Cosmology (MoE), Shanghai 200240, China}
\address[5]{Shanghai Key Laboratory for Particle Physics and Cosmology, Shanghai 200240, China}
\address[2]{School of Nuclear Science and Technology, Lanzhou University, Lanzhou 730000, China}
\address[3]{Key Laboratory of Special Function Materials and Structure Design, Ministry of Education, Lanzhou 730000, China}

\begin{abstract}
Electric quadrupole transition probabilities $B(E2)$ serve as key probe of nuclear shell evolution, yet anomalous $B(E2)$ values in exotic nuclei complicate the identification of new magic numbers. In this letter, employing the configuration-interaction relativistic Hartree-Fock model, we demonstrate that effective charges are sensitive to orbital radii, and this orbital dependence is significantly amplified by the halo structure of valence nucleons. This mechanism is critical for reliably describing $E2$ transitions and understanding the unusual behavior of $B(E2)$ in exotic nuclei. Our calculations predict reduced $B(E2; 2^+_1 \rightarrow 0^+_1)$ values in $^{52,54}\text{Ca}$, signaling the emergence of subshell closures at $N=32$ and 34. Furthermore, the suppressed $B(E2; 7/2^{-}_{1} \rightarrow 11/2^{-}_{1})$ transition in $^{53}\text{Sc}$ underscores the robustness of the $N=32$ new magic number, whereas the enhanced transition strength in $^{55}\text{Sc}$ indicates the rapid erosion of the $N=34$ shell gap with the occupancy of the proton orbital $\pi1f_{7/2}$.
\end{abstract}

\end{frontmatter}

The nuclear shell structure and associated magic numbers, established by Mayer and Jensen through the introduction of strong spin-orbit coupling \cite{mayer1948PR74.235,haxel1949PR75.1766}, have long underpinned our understanding of complex nuclear phenomena and the development of nuclear many-body theories. However, this paradigm undergoes significant modification when extended from stable to exotic nuclei, as evidenced by the emergence of new magic numbers \cite{Ozawa2000PRL84.5493,Thirolf2000PLB485.16,Prisciandaro2001PLB510.17} and the quenching of conventional ones \cite{Simon1999PRL83.496,Motobayashi1995PLB346.9,Bastin2007PRL99.022503}. Such shell evolution provides a unique opportunity to probe the spin-isospin dependence of the nuclear force and benchmark advanced nuclear models, thereby attracting intense interest \cite{otsuka2020RMP92}.

Observables such as the nucleon separation energies and excitation energies of low-lying states are expected to be highly sensitive to the shell gap around the Fermi level. Consequently, shell closures are typically characterized by elevated excitation energies of the first $2^{+}$ state, $E(2^{+}_{1})$, relative to neighboring nuclei. This behavior is evident at $N = 32$ from Ar to Cr \cite{Steppenbeck2015PRL114.252501,Huck1985PRC31.2226,Janssens2002PLB546.55}, and at $N = 34$ in Ar and Ca isotopes \cite{Liu2019PRL122.072502,Steppenbeck2013Nature502.207}. In contrast, the significantly reduced $E(2^{+}_{1})$ values in $^{32}\mathrm{Mg}$ and $^{30}\mathrm{Ne}$ signal the vanishing of the conventional magic number $N = 20$ \cite{Detraz1979PRC19,Yanagisawa2003PLB566}. Recent mass measurements have corroborated the doubly magic nature of $^{52}\mathrm{Ca}$ and $^{54}\mathrm{Ca}$ \cite{Wienholtz2013Nature498.346,Michimasa2018PRL121.022506}, and further indicate the quenching of the $N = 34$ subshell closure above $Z = 20$ \cite{Leistenschneider2021PRL126,Porter2022PRC106,Iimura2023PRL130}, alongside the erosion of the $N = 20$ shell gap in Mg isotopes \cite{Lykiardopoulou2025PRL134}.

Nuclear transition probabilities and charge radii, governed by nucleonic density distributions and single-particle occupancies, serve as complementary probes of shell structure. The enhanced $B(E2; 2^{+}_{1}\rightarrow 0^{+}_{1})$ values in $^{32}\mathrm{Mg}$ and $^{30}\mathrm{Ne}$ unambiguously signal strong quadrupole collectivity and the breakdown of the $N=20$ magic number \cite{Motobayashi1995PLB346.9,Doornenbal2016PRC93}, while the abrupt increase in charge radii across $^{30}\mathrm{Mg}$ delineates the boundary of the $N=20$ island of inversion \cite{Yordanov2012PRL108.042504}. However, an anomalous decrease in $B(E2)$ has been observed from $^{48}\mathrm{Ca}$ to $^{50}\mathrm{Ca}$ \cite{Valiente2009PRL102}, accompanied by unexpectedly large orbital radii of $\nu2p_{3/2}$ and matter radii beyond $N = 28$ shell closure in Ca \cite{Garcia2016NP12} and K isotopes \cite{Koszorus2021NP17}. Such anomalous behaviors complicate the assessment of shell evolution in exotic nuclei via $B(E2)$ and radii systematics, and pose significant challenges to modern nuclear many-body models.

The configuration interaction shell model, with a valence-space tailored effective Hamiltonian, provides a powerful tool for investigating the properties of low-lying states~\cite{Caurier2005RMP77}. Within this truncated model space, effective charges are routinely introduced to account for core-polarization effects in the evaluation of $B(E2)$ strengths \cite{Navratil1997PRC55}, yet their precise values and orbital dependence remain subject to intense debate, particularly in exotic nuclei. For instance, while the widely-used KB3G and GXPF1A effective Hamiltonians yield satisfactory descriptions of binding energies and low-lying spectra for neutron-rich $pf$-shell nuclei, they fail to capture the $B(E2)$ systematics near the new magic numbers $N=32$ and $34$. This failure persists whether using standard effective charges ($e_{\pi}=1.5e$, $e_{\nu}=0.5e$)~\cite{Dinca2005PRC71.041302, Bueger2005PLB622.29} or values deduced from highly excited states in $A=51$ mirror nuclei ($e_{\pi}=1.15e$, $e_{\nu}=0.8e$). More recently, to explain the $B(E2)$ values for neutron-rich ($N>28$) Ca and Sc isotopes within the UFP-CA effective Hamiltonian, newly fitted effective charges of $e_{\pi}=1.30(8)$ and $e_{\nu}=0.452(7)$ were proposed~\cite{Ogunbeku2025PRL135}. These values are consistent with empirical effective charges adjusted within the $sd$ and $sdpf$ shells~\cite{Dufour1996PRC54.1641}, which exhibit very weak orbital and isospin dependence.

It should be emphasized that effective charges are governed by the specific nuclear structure, and their orbital dependence or local "universality" are expected to stem from a robust microscopic origin. In this Letter, we calculate the effective charges using the Tamm-Dancoff approximation (TDA). It is found that effective charges are highly sensitive to the valence orbital radii, and that the values are modulated significantly by halo structures in the exotic nuclei. Based on this, we provide a self-consistent explanation for the observed anomalous $B(E2)$ values in neutron-rich Ca isotopes and reveal crucial signatures associated with the quenching or emergence of the $N=32$ and $34$ magic numbers across the Ca and Sc isotopic chains.

To obtain the low-lying spectra and transition probabilities consistently, the configuration-interaction relativistic Hartree-Fock (CI-RHF) model is utilized in this work \cite{Liu2025CPC49}. Within the CI-RHF framework, the single-particle states are determined by initial relativistic Hartree-Fock calculations, which capture the halo nature of loosely bound orbitals and the effects of changing nuclear mean-field. Employing the folded-diagram approach and considering core-polarization corrections \cite{Hjorth1992AP213}, the effective Hamiltonians for specific model space are derived from a universal phenomenological Lagrangian, in which density-dependent meson-nucleon coupling strengths are introduced to account for complex in-medium effects in finite nuclear systems \cite{Long2006PLB639.242}. In the evaluation of $\hat{Q}$-box, we only consider the second order corrections to the effective interactions, and the contributions from two-particle and two-hole excitations are included solely to renormalize the pairing interactions. Moreover, the single-particle energy cutoff is set to 80~MeV to contain sufficient intermediate particle-hole excitations.

The reduced electric quadrupole transition probability from the initial state $\Psi_{i}$ to the final state $\Psi_{f}$ can be expressed as,
\begin{equation}
    B(E2; \Psi_{i}\rightarrow\Psi_{f})
    = \frac{1}{2J_{i}+1}|\langle \Psi_{f}||\hat{Q}^{\mathrm{eff}}_{2}||\Psi_{i}\rangle|^{2},
\end{equation}
where the effective electric quadrupole operator $\hat{Q}^{\mathrm{eff}}_{2}$ is introduced to account for the transitions outside the model space. It should be noted that although the bare quadrupole operator only contains one-body term, the effective operator, derived from the many-body renormalization, contains two-body and many-body contributions, which is difficult to evaluate exactly in practical calculations. In present work, the electric quadrupole transition operator $\hat{Q}^{\mathrm{eff}}_{2}$ is truncated to one-body term, and the effective charges $e^{\mathrm{eff}}_{ab}$ are introduced to incorporate core-polarization effects,
\begin{equation}\label{effe}
    \hat{Q}^{\mathrm{eff}}_{2k}
    = \sum_{ab}e^{\mathrm{eff}}_{ab}
    \langle a|r^2Y_{2k}|b\rangle
    c^{\dagger}_{a}c_{b}.
\end{equation}

The effective charges $e^{\mathrm{eff}}_{ab}$ are evaluated microscopically by resumming the particle-hole bubble diagrams to all orders within the Tamm-Dancoff approximation (TDA), based on the self-consistent relativistic Hartree-Fock single-particle basis, which captures the long tails characteristic of weakly bound orbitals. With the orbital- and isospin-dependent TDA effective charges, the CI-RHF calculations provide a reliable description for the $B(E2)$ strengths of Ne isotopes from stability to the neutron drip line, and predict the coexistence of a nearly spherical excited $0^{+}_{2}$ state and a deformed ground state in $^{30}\mathrm{Ne}$ \cite{Liu2026E2eff}.

\begin{figure}[htbp]\setlength{\abovecaptionskip}{0.0em}
  \centering
  \includegraphics[width=0.95\linewidth]{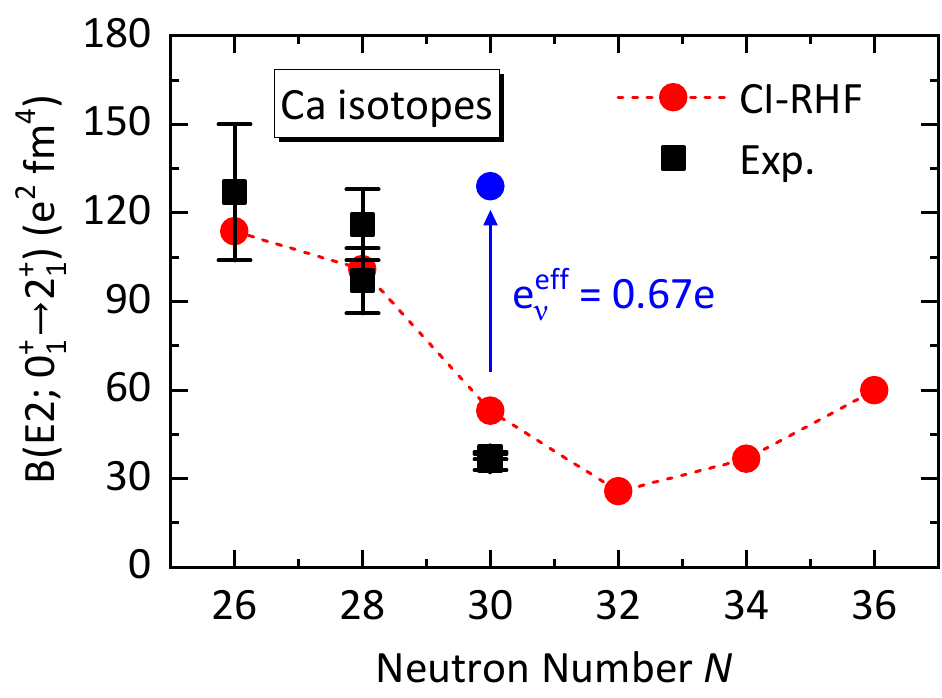}
  \caption{Electric quadrupole reduced transition probabilities $B(E2; 2^{+}_{1}\rightarrow0^{+}_{1})$ for neutron-rich calcium isotopes. Red filled circles connected by a dashed line denote CI-RHF calculations with the PKA1 Lagrangian using self-consistent effective charges; the blue filled circle shows the result with a fixed neutron effective charge $e^{\mathrm{eff}}_{\nu} = 0.67e$. Experimental data (black filled squares with error bars) are taken from Ref.~\cite{Valiente2009PRL102,NNDC,Pritychenko2016ADT107}.}\label{fig:be2_ca}
\end{figure}

Using the CI-RHF model with the PKA1 Lagrangian~\cite{Long2007PRC76.034314}, we calculate $B(E2; 0^{+}_{1}\rightarrow 2^{+}_{1})$ values for even-even $^{46\text{--}56}$Ca isotopes within the $pf$-shell valence space. As shown in Fig.~\ref{fig:be2_ca}, the CI-RHF results with self-consistent effective charges (filled red circles connected by a dashed line) reproduce the experimental $B(E2)$ systematics from $^{46}$Ca to $^{50}$Ca, notably capturing the pronounced drop between $^{48}$Ca and $^{50}$Ca. The calculated $B(E2; (1/2)^{-}_{1} \rightarrow (5/2)^{-}_{1})$ value for $^{55}$Ca is 5.16~$e^{2}\mathrm{fm}^{4}$, also in good agreement with the recent measurement of $5.2 \pm 2.2~e^{2}\mathrm{fm}^{4}$. This success contrasts with previous shell-model studies that employed fixed isoscalar effective charges ($e^{\mathrm{eff}}_{\pi} = 1.5e$, $e^{\mathrm{eff}}_{\nu} = 0.5e$) to describe the anomalously low $B(E2)$ in $^{50}$Ca \cite{Valiente2009PRL102,Riley2014PRC90}, neglecting the isovector quadrupole resonance induced by its significant neutron skin \cite{Tanaka2020PRL124.102501}.

\begin{figure}[htbp]\setlength{\abovecaptionskip}{0.0em}
  \centering
  \includegraphics[width=0.90\linewidth]{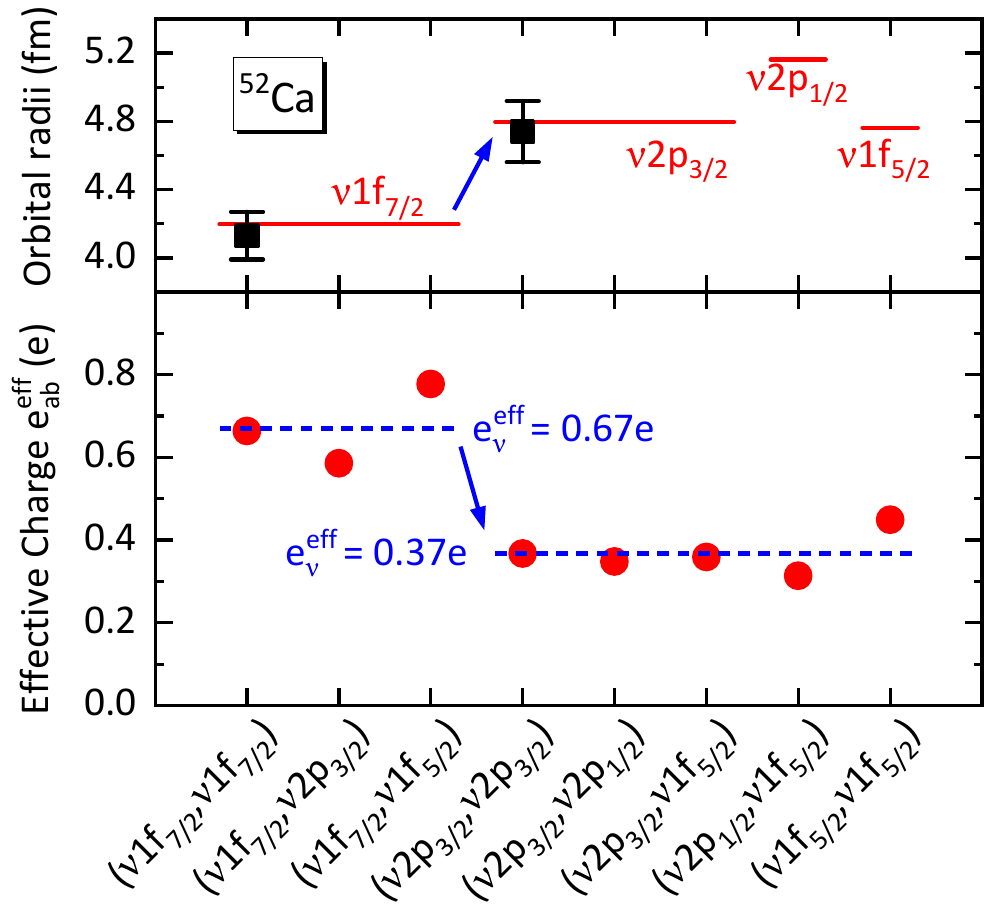}
  \caption{Calculated neutron effective charges $e^{\mathrm{eff}}_{ab}$ and root-mean-square (rms) radii of the corresponding initial states $a$ in $^{52}$Ca. The horizontal dashed lines indicate the average values of the effective charges. Experimental rms radii for the $\nu1f_{7/2}$ and $\nu2p_{3/2}$ orbitals are denoted by filled squares with error bars, with data taken from Ref.~\cite{Enciu2022PRL129}. See text for further details.}\label{fig:ec_ca}
\end{figure}

It is noteworthy that, according to the CI-RHF calculations, the $B(E2)$ values are generally reduced for isotopes beyond $^{48}$Ca rather than solely in $^{50}$Ca, which indicates weakened core-polarization effects for valence orbitals above the $N=28$ shell closure. The lower panel of Fig.~\ref{fig:ec_ca} further displays the consistently calculated effective charges $e^{\mathrm{eff}}_{ab}$ for different valence neutron orbitals. It is found that the average effective charge associated with the $\nu1f_{7/2}$ orbital is $e^{\mathrm{eff}}_{\nu} = 0.67e$, whereas those for orbitals above the $N = 28$ shell closure, namely $\nu2p_{3/2}$, $\nu2p_{1/2}$, and $\nu1f_{5/2}$, are significantly smaller, with an average value of $e^{\mathrm{eff}}_{\nu} = 0.37e$. To assess the impact of this pronounced orbital dependence, we recalculated the $B(E2; 2^+_1 \to 0^+_1)$ value for $^{50}$Ca using a fixed average effective charge of $e^{\mathrm{eff}}_{\nu} = 0.67e$, represented by the blue filled circle in Fig.~\ref{fig:be2_ca}. A comparison with the self-consistent results reveals that the pronounced suppression of the $B(E2; 2^+_1 \to 0^+_1)$ strength from $^{48}$Ca to $^{50}$Ca stems chiefly from the weaker polarization charge associated with the $\nu2p_{3/2}$ orbital, since the $B(E2)$ strength in $^{48}$Ca is dominated by the quadrupole transition from $\nu1f_{7/2}$ to $\nu2p_{3/2}$ orbital, whereas that in $^{50}$Ca arises primarily from the transition within the $\nu2p_{3/2}$ orbitals.

To further elucidate the origin of the pronounced orbital dependence of neutron polarization charges, we calculate the rms radii of the valence neutron orbitals in $^{52}$Ca. As shown in the upper panel of Fig.~\ref{fig:be2_ca}, the rms radii of valence neutron orbitals above the $N=28$ shell closure are significantly increased. Consistent with the momentum distribution analysis \cite{Enciu2022PRL129}, the $\nu2p_{3/2}$ orbital exhibits a pronounced halo-like character, with an rms radius approximately 0.6\,fm larger than that of the $\nu1f_{7/2}$ orbital. Combined with the lower panel of Fig.~\ref{fig:ec_ca}, it is evident that the neutron polarization charges are highly sensitive to these rms radii. Specifically, compared to the tightly bound $\nu1f_{7/2}$ orbital, the more extended density distributions of the $\nu2p$ and $\nu1f_{5/2}$ orbitals weaken their coupling to the $^{40}$Ca core. This suppression is particularly strong for the $\nu2p$ orbitals due to the presence of a radial node, which reduces the core-polarization effects and leads to the observed decrease in both neutron polarization charges and $B(E2)$ values beyond the $N=28$ shell closure. Notably, the calculated neutron effective charges exhibit relatively weak orbital dependence among the orbitals above the $N=28$ gap, justifying the use of a fixed neutron effective charge for $B(E2)$ evaluations in neutron-rich $pf$-shell nuclei \cite{Ogunbeku2025PRL135}.

By incorporating self-consistent effective charges and explicitly accounting for halo effects in weakly bound systems, the CI-RHF model is expected to provide reliable predictions of $B(E2)$ values for neutron-rich isotopes. As shown in Fig.~\ref{fig:be2_ca}, the calculated $B(E2; 0^+_1 \rightarrow 2^+_1)$ values for $^{52}$Ca and $^{54}$Ca are markedly suppressed relative to those of their open-shell neighbors. The CI-RHF calculations reveal that these $E2$ transitions are dominated by single-particle excitations across the $N=32$ and $34$ shell gaps, which naturally leads to the weak quadrupole collectivity of the yrast band. Consequently, the predicted suppression serves as an important signature of persistent neutron subshell closures at $N=32$ and $34$ in Ca isotopes, corroborating recent mass measurements and low-lying spectroscopic data \cite{Wienholtz2013Nature498.346,Michimasa2018PRL121.022506}.

\begin{figure}[htbp]\setlength{\abovecaptionskip}{0.0em}
  \centering
  \includegraphics[width=0.95\linewidth]{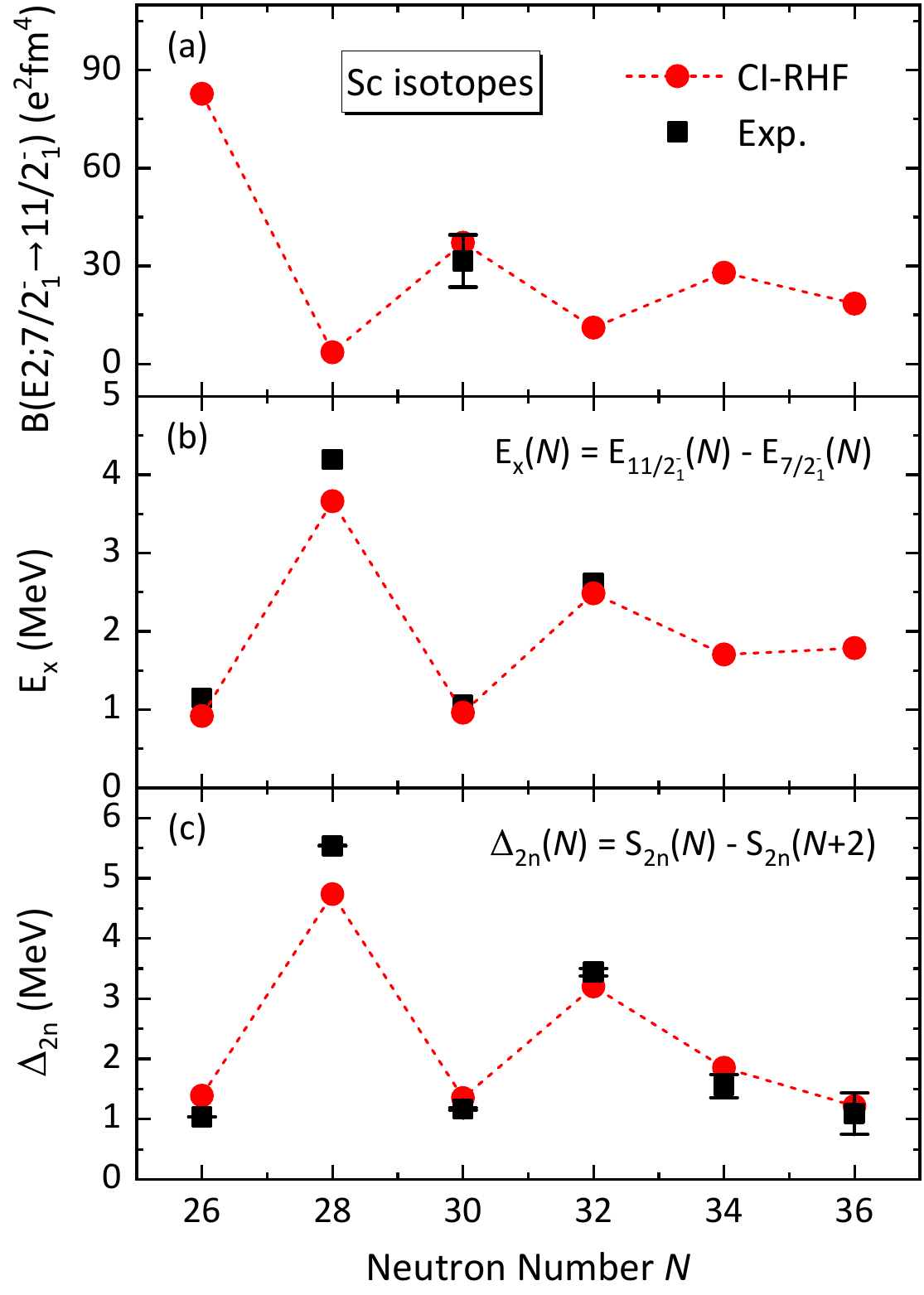}
  \caption{Reduced electric quadrupole transition probabilities $B(E2)$, excitation energies $E_x$, and empirical two-neutron separation energy gaps $\Delta_{2n}$ for Sc isotopes. Red filled circles connected by dashed lines represent CI-RHF calculations with the PKA1 Lagrangian. Experimental data (filled squares with error bars) are taken from Refs.~\cite{Pritychenko2016ADT107,NNDC,Wang2017CPC41.030003}; notably, $\Delta_{2n}$ values derived from recent mass measurements \cite{Leistenschneider2021PRL126,Meisel2020PRC101,Michimasa2020PRL126} are indicated by purple filled squares.}\label{fig:be2_sc}
\end{figure}

Beyond the $Z=20$ shell closure, the $N=34$ shell gap is expected to weaken driven by the proton--neutron monopole interactions. This trend is clearly evident in Ti and heavier elements, where both a reduced $2^+_1$ excitation energy---along with recent high-precision mass measurements---support the quenching of the $N=34$ shell closure. However, the situation in Sc isotopes, which lie between the doubly magic Ca ($Z = 20$) and Ti ($Z = 22$), has remained ambiguous. As shown in Fig. \ref{fig:be2_sc}, recent high-precision mass measurements of neutron-rich Sc isotopes show quantitative differences from AME2016: the new data yield a relatively small two-neutron shell gap at $N=32$ but a larger one at $N=34$. Despite this discrepancy, both datasets indicate a pronounced shell gap at $N=32$ and a rapidly diminishing one at $N=34$---a behavior markedly distinct from that in Ca isotopes. However, neither the widely used phenomenological interactions (KB3G and GXPF1A) nor the ab initio interactions derived by the valence-space in-medium similarity renormalization group approach can simultaneously reproduce the experimental two-neutron shell gap $\Delta_{2n}$ at $N=34$ in both Ca and Sc isotopes.

Motivated by its successful description of the systematics of $B(E2)$ values and nuclear masses in Ca isotopes~\cite{Liu2020PLB806}, the CI-RHF model is employed to investigate the shell structure in Sc isotopes. Despite a slight underestimation of the $N=28$ shell gap, the newly measured two-neutron energy gaps $\Delta_{2n}$ at $N=32$ and $N=34$ are quantitatively reproduced, as shown in Fig.~\ref{fig:be2_sc}(c). To further elucidate the shell evolution from Ca to Sc, we extract key properties of the excited state $11/2^{-}_{1}$, namely its excitation energy $E_x$ and the reduced transition probability $B(E2; 7/2^{-}_{1} \rightarrow 11/2^{-}_{1})$. In a simple shell-model picture without residual interactions, the $11/2^{-}_{1}$ state can be interpreted as arising from the coupling of the proton orbital $\pi 1f_{7/2}$ with the $2^{+}_{1}$ state of the Ca core. As shown in Fig.~\ref{fig:be2_sc}(b), pronounced increases in $E_x$ are observed at $N = 28$ and $N = 32$ relative to neighboring isotopes, in excellent agreement with experiment. These enhancements stem from single-particle cross-shell excitations: specifically, the promotion of neutrons from the $\nu 1f_{7/2}$ to the $\nu 2p_{3/2}$ orbital at $N = 28$, and from $\nu 2p_{3/2}$ to $\nu 2p_{1/2}$ at $N = 32$. Notably, the excitation energies at these neutron numbers are comparable to those of the $2^{+}_{1}$ states in Ca isotopes, underscoring the persistence of the $N = 32$ shell closure beyond $Z = 20$. In stark contrast, the CI-RHF model predicts a marked reduction in $E_x$ at $N = 34$, close to the value in the open-shell nucleus $^{57}$Sc, consistent with the newly mass measurements \cite{Leistenschneider2021PRL126}.

With self-consistent effective charges, the CI-RHF calculations reproduce the experimental $B(E2; 7/2^{-}_{1} \rightarrow 11/2^{-}_{1})$ value in $^{51}$Sc, as shown in Fig.~\ref{fig:be2_sc}(a). Compared with Ca isotopes, the $E2$ transition matrix element $\langle \Psi_{f} \Vert E2 \Vert \Psi_{i} \rangle$ in $^{51}$Sc is significantly larger than that in $^{50}$Ca. This is not surprising, as a portion of valence protons have been excited to the $\pi 2p_{3/2}$ orbital in the $7/2^{-}_{1}$ ground state of open-shell Sc isotopes, contributing substantially to the quadrupole moment. Recently, the $B(E2; 1^{+}_{1} \rightarrow 3^{+}_{1})$ strength in the odd-odd nucleus $^{54}$Sc was extracted from experiment as 23.4$\pm$1.9 e$^{2}\mathrm{fm}^{4}$, and the present calculation yield a reasonable value of 28.5 e$^{2}\mathrm{fm}^{4}$.

Notably, the CI-RHF calculations predict reduced $B(E2)$ strengths at $N=28$ and $N=32$ in Sc isotopes, as shown in Fig. \ref{fig:be2_sc}. The calculated quadrupole density reveal that the single-neutron cross-shell excitation dominates these $E2$ transitions, while the contribution from valence protons is negligible, supporting the closed-shell neutron configuration in $^{49}$Sc and $^{53}$Sc. Due to the occupation of the valence proton orbital $\pi1f_{7/2}$, the neutron quadrupole density associated with the cross-shell excitation from $\nu 1f_{7/2}$ to $\nu 2p_{3/2}$ is significantly diminished in $^{49}$Sc, resulting in a suppressed $B(E2)$ relative to $^{48}$Ca. Moreover, the CI-RHF calculations show that the $B(E2)$ markedly increases at $N = 34$ in Sc isotopes, which are attributed by neutron de-excitation from $\nu1f_{5/2}$ to $\nu2p_{3/2}$ and proton de-excitation from $\pi2p_{3/2}$ to $\pi1f_{7/2}$. Such transitions indicate significant cross-shell excitations in the $7/2^{-}_{1}$ ground state of $^{55}$Sc, driven by the weaken of the $N = 34$ shell gap and strong proton–neutron quadrupole correlations. In fact, the trend of $B(E2; 7/2^{-}_{1} \rightarrow 11/2^{-}_{1})$ in neutron-rich Sc isotopes exhibits a similar isospin evolution to that of $B(E2; 0^{+}_{1} \rightarrow 2^{+}_{1})$ in Ti isotopes---both show pronounced reductions at $N = 28$ and $32$, and a marked enhancement at $N = 34$. Thus, the $B(E2)$ systematics serve as a robust signature of the persistence of the $N = 32$ shell closure and the rapid erosion of the $N = 34$ shell gap in this region.

In summary, by employing the CI-RHF framework with self-consistent effective charges that incorporate all-order core-polarization corrections via the Tamm-Dancoff approximation, we demonstrate that the effective charges are exquisitely sensitive to the radii of valence orbitals. This sensitivity is dramatically amplified by halo effects in weakly bound systems, providing a natural explanation for the observed reduction in $B(E2;0^+_1 \to 2^+_1)$ from $^{48}$Ca to $^{50}$Ca. Our work establishes $B(E2)$ transition strengths as a powerful and direct probe of shell evolution in exotic nuclei, on par with masses and excitation spectra. The predicted suppression of $B(E2)$ values in $^{52,54}$Ca robustly signals the persistence of the $N=32$ and $N=34$ shell closures along the Ca chain. Furthermore, the combined evidence from the suppressed $B(E2;7/2^-_1 \to 11/2^-_1)$ transition in $^{53}$Sc, enhanced two-neutron energy gap, and elevated excitation energies provides compelling support for $N=32$ as a robust magic number beyond $^{52}$Ca. In stark contrast, the predicted enhancement of the $B(E2)$ strength and reduction in excitation energy---complemented by the weak two-neutron shell gap observed in $^{55}$Sc---signal the rapid erosion of the $N=34$ shell structure with increasing proton number.

\section*{Acknowledgement}
This work is partly supported by the National Natural Science Foundation of China under Grant Nos. 12547177, 11675065 and 11875152, the Strategic Priority Research Program of Chinese Academy of Sciences, Grant No. XDB34000000, and Fundamental Research Funds for the Central Universities under Grant No. lzujbky-2019-11.

\bibliographystyle{apsrev4-2}
\bibliography{reference_fixed}

\end{document}